\documentclass[aps,prb,twocolumn,showpacs,groupaddress,floatfix]{revtex4}
\usepackage{graphicx,graphics}
\usepackage{dcolumn}
\usepackage{amsmath,amssymb,amsfonts}
\usepackage{latexsym,verbatim}
\usepackage{bm}
\usepackage{color}
\usepackage{ulem}
\usepackage[breaklinks=true,colorlinks,citecolor=blue,linkcolor=blue,urlcolor=blue]{hyperref}
\def\be{\begin{equation}}
\def\ee{\end{equation}}
\begin{document}
\title{The impact of disorder on Dirac plasmon losses}
\author{Alessandro Principi}
\email{principia@missouri.edu}
\affiliation{Department of Physics and Astronomy, University of Missouri, Columbia, Missouri 65211, USA}	
\author{Giovanni Vignale}
\affiliation{Department of Physics and Astronomy, University of Missouri, Columbia, Missouri 65211, USA}
\author{Matteo Carrega}
\affiliation{NEST, Istituto Nanoscienze-CNR and Scuola Normale Superiore, I-56126 Pisa, Italy}
\author{Marco Polini}
\affiliation{NEST, Istituto Nanoscienze-CNR and Scuola Normale Superiore, I-56126 Pisa, Italy}

\begin{abstract}
Recent scattering-type scanning near-field optical spectroscopy (s-SNOM) experiments on single-layer graphene have reported Dirac plasmon lifetimes that are substantially shorter than the dc transport scattering time $\tau_{\rm tr}$. We highlight that the plasmon lifetime is fundamentally different from $\tau_{\rm tr}$ since it is controlled by the imaginary part of the current-current linear response function at {\it finite} momentum and frequency. We first present the minimal theory of the extrinsic lifetime of Dirac plasmons due to scattering against impurities. We then show that a very reasonable concentration of charged impurities yields a plasmon damping rate which is in good agreement with s-SNOM experimental results.
\end{abstract}
\pacs{73.20.Mf,71.45.Gm,78.67.Wj}
\maketitle

{\it Introduction.---}Graphene plasmonics~\cite{koppens_nanolett_2011,grigorenko_naturephoton_2012} is a rapidly growing branch of research which aims at exploiting the interaction of infrared light with the so-called ``Dirac plasmons" (DPs)~\cite{DiracplasmonsRPA,principi_prb_2009,orlita_njp_2012,abedinpour_prb_2011} for a variety of applications ranging from photodetectors~\cite{freitag_naturecommun_2013,vicarelli_naturemater_2012} to biosensors~\cite{grigorenko_naturemater_2013}. DPs, the self-sustained density oscillations of the two-dimensional (2D) electron liquid in a doped graphene sheet~\cite{novoselov_naturemater_2007,kotov_rmp_2012}, have been studied experimentally by a variety of spectroscopic methods~\cite{grigorenko_naturephoton_2012}. 

Fei {\it et al.}~\cite{fei_nature_2012} and Chen {\it et al.}~\cite{chen_nature_2012} have carried out seminal scattering-type scanning near-field optical microscopy (s-SNOM) experiments in which DPs are launched and imaged in real space. They showed that the plasmon wavelength $\lambda_{\rm p}$ can be $\sim 40$-$60$ times smaller than the illumination wavelength, allowing an extreme concentration of electromagnetic energy, and that DP properties are easily gate tunable. They have also presented an experimental analysis of DP losses. By comparing theory to experimental data, Fei {\it et al.}~\cite{fei_nature_2012} have reported a DP damping rate $\gamma_{\rm p} \simeq 0.08$ (subtracting here the damping due to the complex dielectric constant of the substrate), which is four times larger than that estimated by means of the Drude transport time of their samples. Similarly, Chen {\it et al.}~\cite{chen_nature_2012} showed that theoretical calculations of the local density of optical states compare  well with measurements when rather low values, $\sim 1200~{\rm cm}^2/({\rm V} {\rm s})$, of the sample mobility are used as inputs in the numerics.

The plasmon damping rate (or inverse quality factor), which is a key figure of merit of nanoplasmonics, is defined as~\cite{principi_arxiv_2013}
\begin{equation}\label{eq:dampingrate}
\gamma_{\rm p}(q) \equiv 2\frac{\Gamma_{\rm p}(q)}{\omega_{\rm p}(q)}~, 
\end{equation}
where $\omega_{\rm p}(q)$ is the DP dispersion~\cite{DiracplasmonsRPA,principi_prb_2009,orlita_njp_2012,abedinpour_prb_2011} and $\Gamma_{\rm p}(q)$ is its linewidth~\cite{Giuliani_and_Vignale}. The factor two on the right-hand side of Eq.~(\ref{eq:dampingrate}) has been deliberately added to make direct contact with the results of Ref.~\onlinecite{fei_nature_2012}. The DP damping rate is controlled by several mechanisms, such as electron-electron (e-e), electron-phonon (e-phon), and electron-impurity (e-imp) scattering. The rates associated to the first two scattering mechanism have been theoretically studied in Refs.~\onlinecite{principi_arxiv_2013} and ~\onlinecite{yan_naturephoton_2013}, respectively. To the best of our knowledge, the impact of e-imp scattering on the DP lifetime has not yet been quantitatively analyzed. We believe that this is because the plasmon lifetime associated with e-imp scattering is typically~\cite{koppens_nanolett_2011,fei_nature_2012,chen_nature_2012} believed to be close to the dc Drude transport time $\tau_{\rm tr}$. It is well known~\cite{belitz_prb_1986} that this is not the case and that the plasmon lifetime $\tau_{\rm p}(q)\equiv [2\Gamma_{\rm p}(q)]^{-1}$ can be substantially smaller than $\tau_{\rm tr}$. 

A microscopic understanding of DP losses is central to the success of graphene as a novel platform for nanoplasmonics. As a step towards the elucidation of the mechanisms that contribute to the DP damping rate, in this Rapid Communication we present the simplest theory of the DP plasmon damping rate due to e-imp scattering. Although our theory is general, our numerical calculations focus, for the sake of concreteness, 
on the role of {\it charged} impurities. Long-range Coulomb disorder is indeed the most ``popular"~\cite{dassarma_rmp_2011} candidate for the main scattering mechanism limiting mobility in graphene sheets deposited on substrates like ${\rm SiO}_2$. Other important sources of disorder, such as corrugations~\cite{gibertini_prb_2012} and resonant scatterers~\cite{Katsnelsonbook} can also affect the DP lifetime. In this respect, we note that Yuan {\it et al.}~\cite{yuan_prb_2011} have shown that even a small amount of resonant scatterers such as lattice defects or adsorbates can account for the observed~\cite{li_natphys_2008} background of optical absorption below the single-particle threshold. A comparative study of the impact of various disorder models on the DP lifetime is beyond the scope of the present work.

In this Rapid Communication we demonstrate that neglecting the dependence of the ``memory kernel" $1/\tau(q,\omega)$ on wavevector $q$ and frequency $\omega$---for example by approximating $\tau(q,\omega)$ with $\tau_{\rm tr}$---results in a severe underestimation of disorder-induced DP losses. We find that a very reasonable concentration of charged impurities is enough to explain the experimental findings of Refs.~\onlinecite{fei_nature_2012,chen_nature_2012}. Our results, in combination with those reported in Ref.~\onlinecite{principi_arxiv_2013}, strongly suggest that current s-SNOM experiments~\cite{fei_nature_2012,chen_nature_2012} are dominated by disorder. As suggested in Ref.~\onlinecite{principi_arxiv_2013}, the ``intrinsic" regime where many-body effects dominate DP losses can be reached in suspended graphene sheets or in graphene flakes deposited on hexagonal Boron Nitride.

{\it Theoretical formulation.---}DP losses are quantified by the plasmon damping rate~\cite{footnotecomplex} 
$
\gamma_{\rm p}(q) = {\cal R}\big(q,\omega_{\rm p}(q)\big)
$,
where
\begin{equation} \label{eq:sigma_ratio}
{\cal R}(q,\omega) \equiv \frac{\Re e [\sigma(q, \omega)]}{\Im m [\sigma(q,\omega)]}~,
\end{equation}
and $\omega_{\rm p}(q) = \sqrt{2 {\cal D}_0 q/\epsilon}$ is the long-wavelength plasmon dispersion calculated at the level of the random phase approximation (RPA)~\cite{DiracplasmonsRPA}. Here ${\cal D}_0 = 4 \varepsilon_{\rm F} \sigma_{\rm uni}/\hbar$ is the Drude weight~\cite{grigorenko_naturephoton_2012} of non-interacting massless Dirac fermions (MDFs), $\sigma_{\rm uni} = N_{\rm f} e^2/(16\hbar)$ is the universal conductivity~\cite{grigorenko_naturephoton_2012}, and $N_{\rm f}=4$ is the number of fermion flavors in graphene stemming from spin and valley degrees of freedom. We also introduced $\epsilon = (\epsilon_{\rm air} + \epsilon_{\rm sub})/2$, {\it i.e.} the average of the dielectric constants of the media above ($\epsilon_{\rm air}  = 1$) and below ($\epsilon_{\rm sub}$) the graphene flake. Finally, $\varepsilon_{\rm F} = \hbar v_{\rm F} k_{\rm F}$ is the Fermi energy~\cite{electronholesymmetry}, where $k_{\rm F} = \sqrt{4\pi n/N_{\rm f}}$ is the Fermi wavevector. The Fermi velocity $v_{\rm F}$ is $\sim 10^6~{\rm m}/{\rm s}$.

It is crucial to note that the relation between $\gamma_{\rm p}(q)$ and the conductivity differs from that given in Ref.~\onlinecite{fei_nature_2012}. In deriving a similar relation, the authors of Ref.~\onlinecite{fei_nature_2012} have {\it neglected} the dependence of the conductivity $\sigma(q,\omega)$ on wavevector $q$. This quantity is related to the density-density response function $\chi_{nn}(q,\omega)$ by~\cite{Giuliani_and_Vignale}
$
\sigma(q,\omega) = i e^2 \omega \chi_{nn} (q,\omega)/q^2
$. 
Below we calculate $\Im m[\chi_{nn}(q,\omega)]$ at finite $q$ and $\omega$ to the lowest non-vanishing order ({\it i.e.}  second order) in the e-imp potential.

The homogeneous optical conductivity of non-interacting MDFs in a doped graphene sheet is given by~\cite{abedinpour_prb_2011}
$
\sigma_{\rm c}(q=0,\omega) = i{\cal D}_0[\pi (\omega + i 0^+)]^{-1}
$. 
The previous result, which expresses the intraband contribution to the long-wavelength conductivity, is valid for $\hbar \omega < 2\varepsilon_{\rm F}$ and in the absence of disorder. Note that $\Re e[\sigma_{\rm c}(q=0,\omega)]$ has a delta function peak (the so-called ``Drude peak'') at $\omega=0$, whose strength is given by ${\cal D}_0$.
 
In the presence of weak disorder and in the spirit of the Drude transport theory, it is  natural to define a wavevector- and frequency-dependent scattering time $\tau(q,\omega)$ as follows~\cite{Giuliani_and_Vignale}:
\begin{equation} \label{eq:sigma_omega_def}
\sigma(q,\omega) \equiv \frac{i {\cal D}_0/\pi}{\omega + i \tau^{-1}(q,\omega)}~.
\end{equation}
This expression is valid in the limit of $v_{\rm F} q \ll \omega \ll 2\varepsilon_{\rm F}/\hbar$: this is precisely the region of the $(q,\omega)$ plane where the DP lives~\cite{DiracplasmonsRPA}. The function $\tau(q,\omega)$ has been assumed real. The usual dc Drude transport time is given by
$
\tau_{\rm tr} \equiv \lim_{\omega \to 0} \tau(q=0,\omega)
$. 
As expected, a finite transport time broadens the zero-frequency Drude peak into a Lorentzian. It is easy to prove that, in the weak scattering limit $\omega \tau(q,\omega) \gg 1$, the DP lifetime is given by $\tau\big(q,\omega_{\rm p}(q)\big)$ and that ${\cal R}(q,\omega) = [\omega\tau(q,\omega)]^{-1}$. In the same limit one gets~\cite{Giuliani_and_Vignale}
\begin{equation} \label{eq:lifetime_def}
\frac{1}{\tau(q,\omega)} = -\frac{\pi e^2 \omega^3}{{\cal D}_0 q^2} \Im m[\chi_{nn} (q,\omega)]
~.
\end{equation}
In deriving Eq.~(\ref{eq:lifetime_def}) we have used that
$
\Re e[\chi_{nn} (q,\omega)] \to {\cal D}_0 q^2/(\pi e^2 \omega^2)
$ in the limit $1, v_{\rm F} q \tau(q,\omega) \ll \omega\tau(q,\omega)\ll 2\varepsilon_{\rm F}\tau(q,\omega)/\hbar$.

Following Ref.~\onlinecite{principi_arxiv_2013}, we describe the electron system in a doped graphene sheet in a tight-binding framework which takes into account only the $\pi$ and $\pi^\star$ bands of graphene. We neglect all the other bands. This approach is sufficient to describe graphene at low energies and eliminates spurious problems associated with the short-distance physics of the MDF model~\cite{abedinpour_prb_2011,sabio_prb_2008}. We take the low-energy MDF limit only {\it after} carrying out all the necessary commutators and algebraic steps briefly sketched in the following. 

Our calculations of the role of e-imp scattering on the DP lifetime are based on the following Hamiltonian: 
${\hat {\cal H}} = {\hat {\cal H}}_0 + {\hat {\cal H}}_{\rm ei}$, where ${\hat {\cal H}}_0$ is the aforementioned tight-binding Hamiltonian (see Ref.~\onlinecite{principi_arxiv_2013} for more details). The e-imp Hamiltonian reads 
$
{\hat {\cal H}}_{\rm ei} = {\cal A}_{\rm BZ}^{-1} \sum_{{\bm q}} u_{\bm q} {\hat n}_{\bm q} n^{({\rm i})}_{-{\bm q}}
$, where ${\bm q}$ is restricted inside the first Brillouin zone and ${\cal A}_{\rm BZ}$ is its area. Here ${\hat n}_{\bm q}$ is the electron density operator~\cite{principi_arxiv_2013} and $n^{({\rm i})}_{\bm q} = \sum_{{\bm R}_i} e^{i {\bm q}\cdot {\bm R}_i}$ is the impurity density. The vector ${\bm R}_i$ labels the random position of the $i$-th impurity. Finally, $u_{\bm q}$ is the {\it discrete} Fourier transform of the e-imp potential. We emphasize that e-e interactions, which, for the sake of simplicity, have not been explicitly added to ${\hat {\cal H}}$, play a twofold role: they enable the existence of plasmons~\cite{Giuliani_and_Vignale} and weaken the bare e-imp potential $u_{\bm q}$ through screening. Both effects are taken into account below at the RPA level.

{\it Elimination of the e-imp potential via a canonical transformation.---}We now calculate $\tau(q,\omega)$ as defined in Eq.~(\ref{eq:lifetime_def}) and in the presence of weak disorder. To this aim, we evaluate $\chi_{nn}(q,\omega)$ on the right-hand side of Eq.~(\ref{eq:lifetime_def}) to second order in the strength of e-imp interactions. Within the tight-binding model, the density operator ${\hat n}_{\bm q}$ can be related to the longitudinal component of the current density operator $\hat{\bm j}_{\bm q}$ by the continuity equation (from now on we set $\hbar =1$):  $i\partial_t {\hat n}_{\bm q} =[{\hat {\cal H}}, {\hat n}_{\bm q}] =  -{\bm q} \cdot {\hat {\bm j}}_{\bm q}$. The continuity equation does not show any anomalous commutator~\cite{abedinpour_prb_2011,sabio_prb_2008} and allows us to express~\cite{principi_arxiv_2013,Giuliani_and_Vignale} $\Im m[\chi_{nn}(q,\omega)]$ in terms of the imaginary part of the longitudinal current-current response function $\chi_{\rm L}(q,\omega)$. In view of the low-energy MDF limit and disorder average taken at the end of the calculation (which restore isotropy), without any lack of generality we can take ${\bm q} = q {\hat {\bm x}}$ and arrive at the desired result:
\be \label{eq:continuity_equation}
\Im m[\chi_{nn}(q,\omega)] = \frac{q^2}{\omega^2}\Im m[\chi_{\rm L}(q,\omega)]
~.
\ee
Eq.~(\ref{eq:continuity_equation}) is the usual relation between density-density and longitudinal current-current response functions, which holds for an isotropic, rotationally-invariant electron liquid. 

To proceed further, we adopt the same strategy detailed in Ref.~\onlinecite{principi_arxiv_2013}. We introduce a unitary transformation generated by an Hermitian operator ${\hat F}$ that cancels exactly the e-imp interaction from ${\hat {\cal H}}$, {\it i.e.}
$
{\hat {\cal H}}' = e^{i {\hat F}} {\hat {\cal H}} e^{-i {\hat F}} \equiv {\hat {\cal H}}_0
$. 
This equation can be solved order by order in perturbation theory, by expanding ${\hat F} = \openone + {\hat F}_1 + {\hat F}_2 + \ldots$, where $\openone$ is the identity operator and ${\hat F}_n$ denotes the $n$-th order term in powers of the strength of e-imp interactions. We obtain a chain of equations connecting ${\hat F}_n$ to ${\hat {\cal H}}_{\rm ei}$. As an example, ${\hat F}_1$ solves the equality $[{\hat {\cal H}}_0, i {\hat F}_1] = {\hat {\cal H}}_{\rm ei}$. 

We then calculate the ``rotated'' current operator, which can be expanded in powers of the e-imp interaction as
$
{\bm q}\cdot{\hat {\bm j}}_{\bm q}' = {\bm q}\cdot{\hat {\bm j}}_{\bm q} + {\bm q}\cdot{\hat {\bm j}}_{1,{\bm q}} + {\bm q}\cdot{\hat {\bm j}}_{2,{\bm q}} + \ldots
$, 
where ${\hat {\bm j}}_{n,{\bm q}}$ is of $n$-th order in the e-imp potential $u_{\bm q}$. Note that only the zeroth-order contribution to ${\bm q}\cdot{\hat {\bm j}}_{\bm q}'$ ({\it i.e.} ${\bm q}\cdot{\hat {\bm j}}_{{\bm q}}$) does not break momentum conservation by transferring part of the momentum ${\bm q}$ to the impurity subsystem. Indeed, it can only generate single-pair excitations of total momentum ${\bm q}$ which lie inside the particle-hole continuum. This in turn implies that in the limit $v_{\rm F} q \ll \omega \ll 2\varepsilon_{\rm F}$ the only non-vanishing second-order contribution in the strength of e-imp interactions to $\Im m[\chi_{\rm L}(q,\omega)]$ is $\Im m[\chi_{j_{1,x} j_{1,x}}(q{\hat {\bm x}},\omega)]$. We find
\begin{eqnarray} \label{eq:SM_j_1_element}
{\bm q}\cdot{\hat {\bm j}}_{1,{\bm q}} = [i {\hat F}_1, {\bm q}\cdot{\hat {\bm j}}_{\bm q}] = {\cal A}_{\rm BZ}^{-1} \sum_{{\bm q}'} u_{{\bm q}'}
{\hat \Upsilon}_{{\bm q}, {\bm q}'} n^{({\rm i})}_{-{\bm q}'}
~.
\end{eqnarray}
It is clear from Eq.~(\ref{eq:SM_j_1_element}) that ${\bm q}\cdot{\hat {\bm j}}_{1,{\bm q}}$ breaks momentum conservation, since an amount $-{\bm q}'$ is transferred to impurities. In the limit $v_{\rm F} q \ll \omega \ll 2\varepsilon_{\rm F}$, the operator ${\hat \Upsilon}_{{\bm q}, {\bm q}'}$ reads
\begin{eqnarray} \label{eq:Upsilon_approx}
{\hat \Upsilon}_{{\bm q},{\bm q}'} &=& 
- \sum_{\alpha=x,y} \left\{
\frac{v_{\rm F}}{\omega^2 k_{\rm F}} q'_x q'_\alpha + \left[ \frac{v_{\rm F}}{\omega^2} \frac{{\bm q}\cdot{\bm q}'}{k_{\rm F}} \left( 3 - \frac{q'^2}{2 k_{\rm F}^2} \right) 
\right.
\right.
\nonumber\\
&-&
\left.
\left. \frac{q'^2}{4 v_{\rm F} k_{\rm F}^3} \right] \delta_{\alpha,x}
\right\}
{\hat j}_{{\bm q}+{\bm q}', \alpha}
\nonumber\\
&\equiv&
- \sum_{\alpha=x,y} \Gamma^{({\rm dis})}_\alpha({\bm q},{\bm q}',\omega) {\hat j}_{{\bm q}+{\bm q}', \alpha}
~.
\end{eqnarray}
Taking the low-energy MDF limit we finally obtain
\begin{eqnarray} \label{eq:transporttime_final}
\frac{1}{\tau(q,\omega)} &=& -\frac{\pi e^2 n_{\rm imp} \omega}{{\cal D}_0} \sum_{\alpha,\beta} \int\frac{d^2{\bm q}}{(2\pi)^2} u_{{\bm q}'}^2 \Gamma^{({\rm dis})}_\alpha({\bm q},{\bm q}',\omega)
\nonumber\\
&\times&
\Gamma^{({\rm dis})}_\beta({\bm q},{\bm q}',\omega)\Im m[\chi_{j_\alpha j_\beta}^{(0)} ({\bm q}+{\bm q}',\omega)]~,
\end{eqnarray}
where the average over disorder $\langle n^{({\rm i})}_{\bm q} n^{({\rm i})}_{{\bm q}'}\rangle_{\rm dis}/{\cal A} =  n_{\rm imp} \delta_{{\bm q}+{\bm q}', 0}$ has been taken. Here $n_{\rm imp}$ is the average impurity density and ${\cal A}$ is the sample area. Eq.~(\ref{eq:transporttime_final}) overestimates the effect of disorder on the electronic system. Indeed, when e-e interactions are taken into account, the bare e-imp potential is weakened. We take into account {\it screening} by replacing in Eq.~(\ref{eq:transporttime_final}) the longitudinal and transverse components of the non-interacting current-current response function $\chi_{j_\alpha j_\beta}^{(0)} ({\bm q}+{\bm q}',\omega)$ with the RPA current-current response $\chi_{j_\alpha j_\beta}^{({\rm RPA})} ({\bm q}+{\bm q}',\omega)$. We remind the reader that the transverse RPA current-current response function coincides with the non-interacting one~\cite{Giuliani_and_Vignale}.

{\it Numerical results.---}We now turn to present our main numerical results for $\tau_{\rm p}(q)\equiv \tau(q,\omega_{\rm p}(q))$ as calculated from Eq.~(\ref{eq:transporttime_final}) with $\chi_{j_\alpha j_\beta}^{(0)} ({\bm q}+{\bm q}',\omega) \to \chi_{j_\alpha j_\beta}^{({\rm RPA})} ({\bm q}+{\bm q}',\omega)$ and for ${\cal R}_{\rm p}(q) \equiv {\cal R}(q,\omega_{\rm p}(q))$---see Eq.~(\ref{eq:sigma_ratio}). For the sake of definiteness, we choose $u_{\bm q}$ to be the long-range potential generated by impurities of unitary charge located on the graphene sheet, {\it i.e.} $u_{\bm q} = 2\pi e^2/(\epsilon q)$. The impurity density $n_{\rm imp}$ is obtained by making sure that the calculated transport time $\tau_{\rm tr}$ equals the experimental value given in Ref.~\onlinecite{fei_nature_2012}, {\it i.e.} $\tau_{\rm exp} = 260~{\rm fs}$, corresponding to a mobility $\mu \sim 8.000~{\rm cm}^2/({\rm V} {\rm s})$, at a carrier density $n=8.0 \times 10^{12}~{\rm cm}^{-2}$. We remind the reader that in this experiment $\epsilon = 2.52$. This constraint is satisfied~\cite{exponential} with an impurity concentration $n_{\rm imp} \simeq 5.8 \times 10^{11}~{\rm cm}^{-2}$.

\begin{figure}[t!]
\begin{center}
\tabcolsep=0cm
\begin{tabular}{c}
\includegraphics[width=1.0\linewidth]{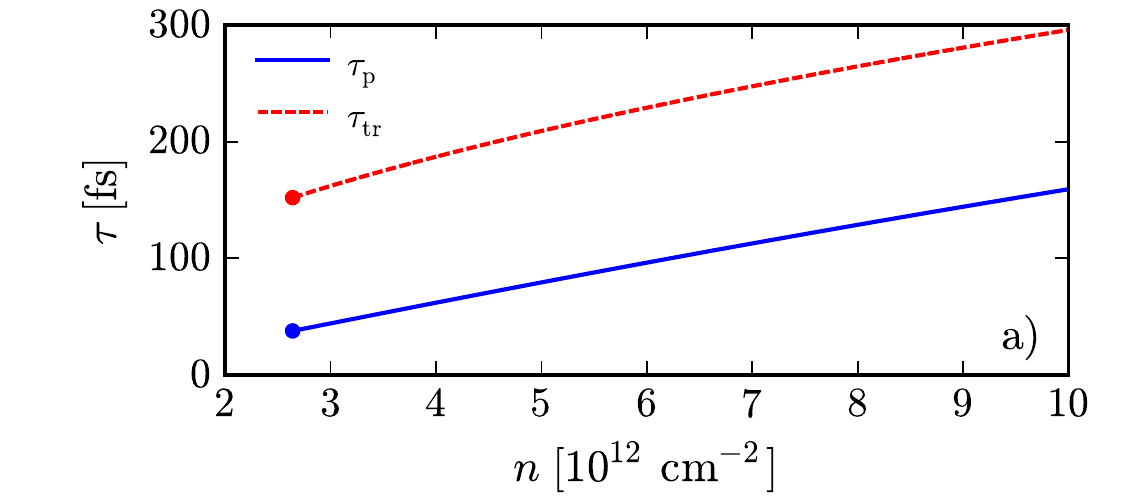}\\
\includegraphics[width=1.0\linewidth]{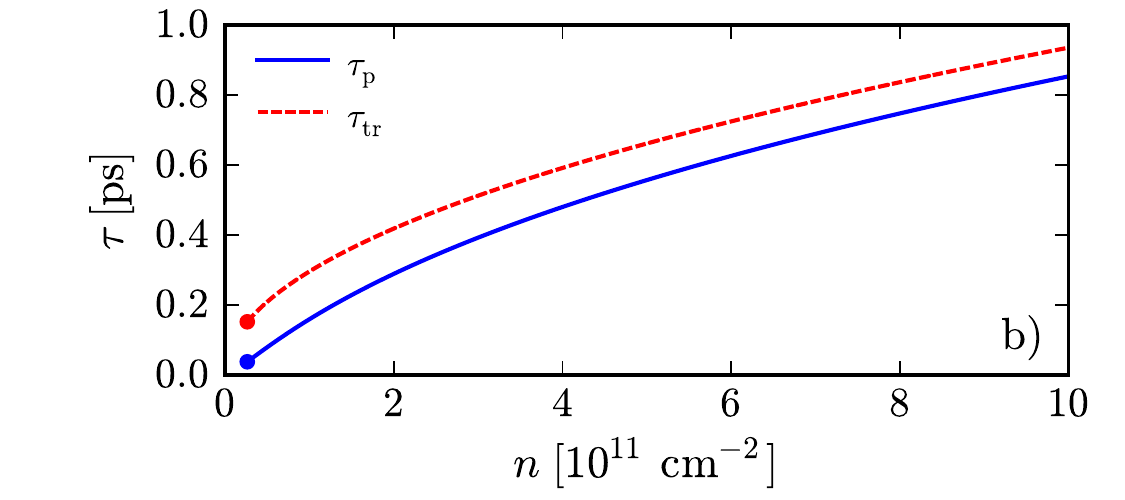}
\end{tabular}
\caption{(Color online) The disorder-induced DP lifetime $\tau_{\rm p}(q_{\rm p})$ and the transport time $\tau_{\rm tr}$ are plotted as functions of the electron density $n$ and for a fixed photon energy $\hbar\omega_{\rm ph}$ and impurity concentration $n_{\rm imp}$. The solid line refers to $\tau_{\rm p}(q_{\rm p})$, while the dashed line refers to $\tau_{\rm tr}$. Different panels refer to different values of the photon energy and impurity concentration: in panel a) we have set $\hbar \omega_{\rm ph} = 112~{\rm meV}$, corresponding to mid-infrared plasmons, and $n_{\rm imp}=5.8 \times 10^{11}~{\rm cm}^{-2}$; in panel b) $\hbar\omega_{\rm ph} = 11.2~{\rm meV}$, corresponding to Terahertz plasmons, and $n_{\rm imp} = 5.8 \times 10^{10}~{\rm cm}^{-2}$. Note the difference in the scales of horizontal and vertical axes between the two panels. In both panels we have set $\epsilon=2.52$, corresponding~\cite{fei_nature_2012} to graphene on ${\rm SiO}_2$.
\label{fig:one}}
\end{center}
\end{figure}
\begin{figure}[t!]
\begin{center}
\tabcolsep=0cm
\begin{tabular}{c}
\includegraphics[width=1.0\linewidth]{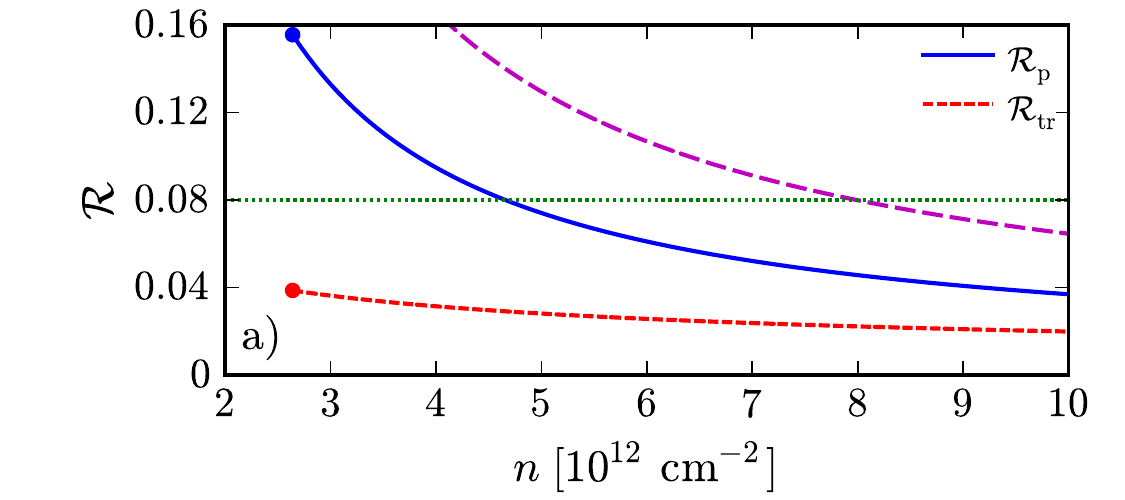}\\
\includegraphics[width=1.0\linewidth]{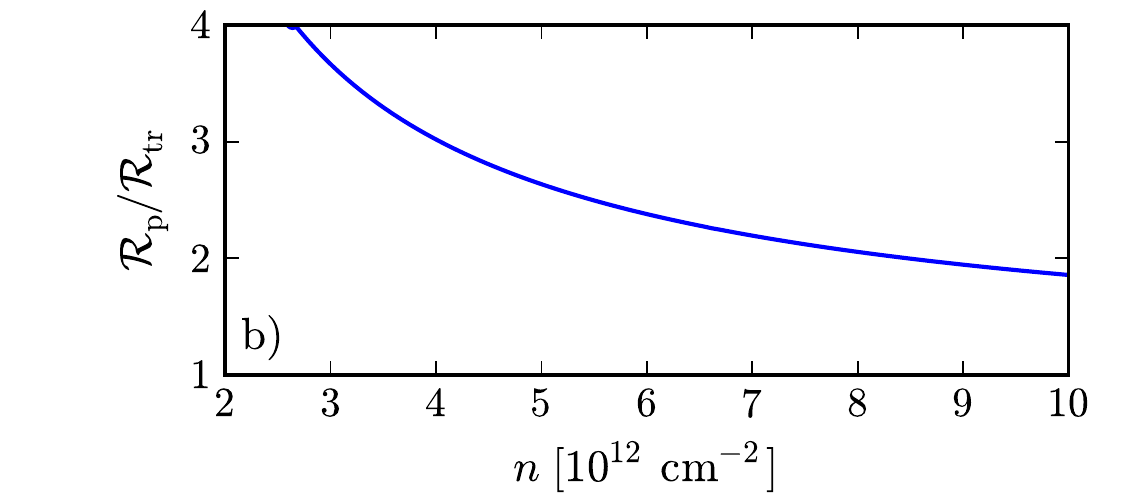}
\end{tabular}
\caption{(Color online) The quantities ${\cal R}_{\rm p}(q_{\rm p})$ and ${\cal R}_{\rm tr}$ are plotted as functions of $n$ and for a fixed $\hbar\omega_{\rm ph}$ and $n_{\rm imp}$. In panel a) the solid (short-dashed) line refers to ${\cal R}_{\rm p}(q_{\rm p})$ (${\cal R}_{\rm tr}$) calculated for $\hbar\omega_{\rm ph} = 112~{\rm meV}$ and $n_{\rm imp} = 5.8\times 10^{11}~{\rm cm}^{-2}$. For the sake of comparison, we show the density-independent experimental value ${\cal R}_{\rm exp} = 0.08$ (dotted line) and ${\cal R}_{\rm p}(q_{\rm p})$ (long-dashed line) calculated for a larger impurity concentration, $n_{\rm imp} = 1.0 \times 10^{12}~{\rm cm}^{-2}$, which matches ${\cal R}_{\rm exp}$ at $n=8.0 \times 10^{12}~{\rm cm}^{-2}$. Panel b) illustrates the ratio ${\cal R}_{\rm p}(q_{\rm p})/{\cal R}_{\rm tr}$ as a function of  $n$ for $\hbar\omega_{\rm ph} = 112~{\rm meV}$. Note that this ratio is independent of the impurity concentration and always larger than unity in the explored range of densities.\label{fig:two}}
\end{center}
\end{figure}

In Fig.~\ref{fig:one}a) we plot the DP lifetime $\tau_{\rm p} (q_{\rm p})$, where the plasmon wavevector is given by~\cite{fei_nature_2012,principi_arxiv_2013} $q_{\rm p}/k_{\rm F} = [\hbar v_{\rm F} \epsilon/(2 e^2)](\hbar \omega_{\rm ph}/\varepsilon_{\rm F})^2$. The photon energy $\hbar \omega_{\rm ph}$ is kept fixed, {\it i.e.} $\hbar \omega_{\rm ph}=112~{\rm meV}$, as in Ref.~\onlinecite{fei_nature_2012}. As density decreases $q_{\rm p}/k_{\rm F}$ increases: filled circles in Fig.~\ref{fig:one} refer to the value of doping such that $(q_{\rm p}/k_{\rm F})_{\rm max} = 0.2$. From this figure we clearly see that the disorder-induced DP lifetime is of the order of $100~{\rm fs}$ for mid-infrared plasmons. As a comparison, we plot also the calculated transport time $\tau_{\rm tr}$, which is clearly larger than $\tau_{\rm p}$. Identifying $\tau_{\rm p}(q_{\rm p})$ with $\tau_{\rm tr}$ leads to an error of factor $\sim 2-3$ in the experimentally relevant range of densities. In Fig.~\ref{fig:one}b) we show our predictions for the DP lifetime for a photon energy of $\hbar \omega_{\rm ph} = 11.2~{\rm meV}$ (Terahertz plasmons). In this case we have fixed $n_{\rm imp} = 5.8\times 10^{10}~{\rm cm}^{-2}$.

In Fig.~\ref{fig:two} we plot the quantity ${\cal R}_{\rm p}(q)$ evaluated at $q = q_{\rm p}$ (at fixed $\omega = \omega_{\rm ph}$). In the same figure we have also plotted the dc value defined as ${\cal R}_{\rm tr} \equiv \lim_{\omega\to 0} {\cal R}(q=0,\omega)$. This figure refers to a photon frequency in the mid-infrared. From panel a), we note that, in the range of densities explored in Refs.~\onlinecite{fei_nature_2012,chen_nature_2012}, the dependence of ${\cal R}_{\rm p}$ on doping is weak. This is in perfect agreement with the findings of Refs.~\onlinecite{fei_nature_2012,chen_nature_2012}. Note that even at carrier densities as large as $10^{13}~{\rm cm}^{-2}$, ${\cal R}_{\rm p}$ is a factor of two larger than ${\cal R}_{\rm tr}$---see panel b). For $n=8.0 \times 10^{12}~{\rm cm}^{-2}$ we find ${\cal R}_{\rm tr} \simeq 0.02$ and ${\cal R}_{\rm p} \simeq 0.05$. The experimentally measured damping rate is represented by a density-independent value~\cite{fei_nature_2012}, ${\cal R}_{\rm exp}=0.08$, which is matched (at $n=8.0 \times 10^{12}~{\rm cm}^{-2}$) by ${\cal R}_{\rm p}$ calculated for an impurity concentration of $1.0 \times 10^{12}~{\rm cm}^{-2}$. Note also that, since the memory kernel $1/\tau(q,\omega)$ scales linearly with  impurity concentration, the ratio ${\cal R}_{\rm p}/{\cal R}_{\rm tr}$ is independent of $n_{\rm imp}$.

{\it Summary and discussion.---}In summary, we have presented a  theory of disorder-induced Dirac plasmon losses. We have carried out numerical calculations for a specific disorder model, {\it i.e.} charged impurities located on graphene. We have shown that the plasmon lifetime is substantially shorter than the dc Drude transport time, even at high carrier densities. The calculated damping rate qualitatively agrees with the experimental findings of Refs.~\onlinecite{fei_nature_2012,chen_nature_2012} and differs by less than a factor of two with respect to the measured value. We stress that the damping rate calculated on the basis of the dc Drude transport time is a factor of four smaller than the measured value.

We stress that the remaining discrepancy between theory and experiments may stem from a number of issues. First, the disorder model we have used (charged impurities) may not be sufficient. Other important sources of extrinsic scattering, such as corrugations, resonant scatterers, and disorder at the edges, may well explain the difference. This remains to be studied. We highlight that scattering of electrons from optical phonons in the ${\rm SiO}_2$ substrate gives a damping rate~\cite{low_private} $\simeq 0.02$, for $\hbar\omega_{\rm p} = 112~{\rm meV}$ and $n=8\times 10^{12}~{\rm cm}^{-1}$. This number, added to the disorder-induced damping rate at the same photon energy and carrier density, gives $\gamma_{\rm p} \simeq 0.07$, in very good agreement with the experimental result. Second, our numerical results heavily rely on an input parameter: the mobility of the samples employed in Ref.~\onlinecite{fei_nature_2012,chen_nature_2012}. This quantity has not been directly measured in Refs.~\onlinecite{fei_nature_2012,chen_nature_2012} but has been inferred from previous measurements on similarly prepared samples. A concentration of charged impurities equal to $n_{\rm imp} = 1.0 \times 10^{12}~{\rm cm}^{-2}$, corresponding to a mobility $\mu \sim 4.000~{\rm cm}^2/({\rm V} {\rm s})$, gives ${\cal R}_{\rm p}  = 0.08$, in perfect agreement with the measured value---see Fig.~\ref{fig:two}a).

Our calculations strongly suggest that plasmon losses in current graphene samples~\cite{fei_nature_2012,chen_nature_2012} are dominated by disorder and that there is plenty of room to increase the sample purity to reach the intrinsic regime of ultra long Dirac plasmon lifetimes~\cite{principi_arxiv_2013}.

{\it Acknowledgements.---}We thank Frank Koppens and Tony Low for useful discussions. A.P. and G.V. were supported by the BES Grant DE-FG02-05ER46203. M.C. and M.P. acknowledge support by MIUR through the program ``FIRB - Futuro in Ricerca 2010" - Project PLASMOGRAPH (Grant No. RBFR10M5BT).

\begin{thebibliography}{77}
\bibitem{electronholesymmetry}
	We discuss electron doping for the sake of definiteness: for weak disorder,
	$\gamma_{\rm p}(q)$ is a particle-hole symmetric quantity.
\end{thebibliography}
\end{document}